%% file: ms.tex
\documentclass[oneside, onecolumn]{article}

\usepackage[utf8]{inputenc}
\usepackage[numbers]{natbib}
\usepackage[justification=centering]{subfig}
\usepackage[nolist,nohyperlinks]{acronym}
\usepackage{color}
\usepackage{booktabs}
\usepackage{verbatim}
\usepackage{amsmath}
\usepackage{amssymb}
\usepackage{amsthm}
\usepackage{mathtools}
\usepackage{datetime}
\usepackage{graphicx}
\usepackage{enumerate}
\usepackage{bbm}  
\usepackage{url}
\usepackage{hyperref}  
\usepackage[margin=1.0in]{geometry}

\newcommand{\todo}[1]{}
\newcommand{\comments}[1]{}
\newcommand{\R}{\mathbb{R}}

\renewcommand{\(}{\left(}
\renewcommand{\)}{\right)}
\newcommand{\<}{\left\langle}
\renewcommand{\>}{\right\rangle}

\renewcommand{\hat}{\widehat}
\renewcommand{\epsilon}{\varepsilon}
\renewcommand{\phi}{\varphi}
\newcommand{\id}{i}

\makeatletter
\def\unmarkedfootnote{\gdef\@thefnmark{}\@footnotetext}
\makeatother

\begin{document}

\title{\huge Fusion of Mobile Device Signal Data Attributes \\ Enables Multi-Protocol Entity Resolution and \\ Enhanced Large-Scale Tracking}
\author{Brian Thompson, Dave Cedel, Jeremy Martin, Peter Ryan, Sarah Kern}
\date{The MITRE Corporation, 7515 Colshire Drive, McLean, VA 22102 \\ \{bthompson, dcedel, jbmartin, peterryan, skern\}@mitre.org}
\unmarkedfootnote{\textcopyright The MITRE Corporation. ALL RIGHTS RESERVED. Approved for public release. Distribution unlimited 18-04590-5.}

\maketitle

\begin{abstract}
{Use of persistent identifiers in wireless communication protocols is a known privacy concern as they can be used to track the location of mobile devices. Furthermore, inherent structure in the assignment of hardware identifiers as well as upper-layer network protocol data attributes can leak additional device information. We introduce SEXTANT, a computational framework that combines improvements on previously published device identification techniques with novel spatio-temporal correlation algorithms to perform multi-protocol entity resolution, enabling large-scale tracking of mobile devices across protocol domains. Experiments using simulated data representing Las Vegas residents and visitors over a 30-day period, consisting of about 300,000 multi-protocol mobile devices generating over 200 million sensor observations, demonstrate SEXTANT's ability to perform effectively at scale while being robust to data heterogeneity, sparsity, and noise, highlighting the urgent need for the adoption of new standards to protect the privacy of mobile device users.}
\end{abstract}

\input{10-introduction}  
\input{20-related}  
\input{30-methods}  
\input{41-evaluation}

\input{50-conclusions}

\section*{Acknowledgments}
\label{sec:acknowledgments}

Views and conclusions are those of the authors and should not be interpreted as representing the official policies or position of the U.S. government.  The authors' affiliation with The MITRE Corporation is provided for identification purposes only, and is not intended to convey or imply MITRE's concurrence with, or support for, the positions, opinions or viewpoints expressed by the authors.

\newpage
\input{ms.bbl}

\newpage
\appendix

\input{60-appendix-convexhull}
\input{61-appendix-parameters}

\begin{acronym}
  \acro{GSM}{Global System for Mobile Communications}
  \acro{GSMA}{GSM Association}
  \acro{GB}{gigabyte}
  \acro{ID}{identifier}
  \acro{IEEE}{Institute of Electrical and Electronics Engineers}
  \acro{IMEI}{International Mobile Equipment Identity}
  \acro{IMSI}{International Mobile Subscriber Identity}
  \acro{IRB}{Institutional Review Board}
  \acro{MAC}{Media Access Control}
  \acro{OS}{Operating System}
  \acro{OUI}{Organizationally Unique Identifier}
  \acro{PII}{Personally Identifiable Information}
  \acro{TMSI}{Temporary Mobile Subscriber Identity}
  \acro{TAC}{Type Allocation Code}
\end{acronym}

\end{document}

%% file: 10-introduction.tex
\section{Introduction}
\label{sec:intro}

\subsection{Motivation}
\label{sec:intro-motiv}

The increasing availability of spatio-temporal data is motivating the development of scalable and efficient algorithms to better leverage that data. Numerous applications across marketing, publicity, social media, tourism, urban planning, and social services rely on data indicating people's locations or mobility patterns~\citep{kontokosta2017urban, nunes2017beanstalk}. Such applications are often limited by insufficient or noisy data.

Simultaneously, we are seeing a recently heightened demand for user privacy, illustrated in practice by the wide-reaching actions of the European Union's General Data Protection Regulation (GDPR)~\citep{eugdprportal}. This mandate reaches across commercial industry, academia, and private business in order to ensure effective attention to user privacy. Achieving that goal, however, is not easy. The ubiquity of mobile devices coupled with a variety of mobile communications platforms complicates the effective implementation of sound privacy. Privacy concerns, specifically regarding the ability to track individual mobile users, are richly described in academic research and are commonly detailed in proactive news articles. Mobile devices, even when not actively utilized by a mobile user, constantly transmit control, management, and data frames, often unbeknownst to the user~\citep{electronicfrontierfoundation, kohno2005remote, wachs2017push, foppe2018exploiting}. These messages contain protocol-specific hardware identifiers that are transmitted in plaintext and are trivially retrieved. The use of these identifiers as tracking mechanisms has been well documented~\citep{cunche2014know, russian, martin2013correlating, martin2016decomposition}. Researchers have described that operating systems designed to curtail such tracking vulnerabilities often leave the user exposed due to implementation design flaws~\citep{vanhoef2016mac, martin2017study, hong2018guti}.

The privacy issues surrounding the leak of individuals' spatio-temporal information is inherently tied to the ability to link devices by the correlation of permanent hardware identifiers or through defeating randomization practices. Until appropriate countermeasures are widely adopted and securely implemented, spatio-temporal algorithms for device correlation remain a viable privacy concern, magnified by the ability to efficiently process large data sets at scale.

Large-scale spatio-temporal data can thus both provide significant societal benefits and pose a significant privacy risk. In introducing our novel computational framework and algorithms, we aim to support the advancement of legitimate services complying with standardized regulations, as well as to call out the inherent privacy risk with current mobile communication implementations and echo the support for the use of temporary identifiers in future incremental design changes.



\subsection{Background}
\label{sec:intro-background}
Wireless frames may contain meta-data including a layer-2 hardware \emph{\ac{ID}} such as a \ac{MAC} address or \ac{IMEI}. In particular, we describe here some of the device-related meta-data for two common protocols used for wireless communications: \ac{GSM} and WiFi/802.11.

By the nature of the allocation process of hardware identifiers these protocols contain a number of inherent data or information leakages. Firstly, the identifiers are intended to be globally unique static identifiers, which has been universally documented as a privacy and tracking concern. While efforts to implement randomization for \ac{MAC} addresses has been implemented in both iOS and some Android devices it has been shown to be flawed and defeatable~\citep{vanhoef2016mac, martin2017study, hong2018guti}.

Furthermore, little use of randomized addressing has been implemented or adopted for permanent connections where data frames still rely on the globally unique permanent identifiers. An exception to this policy and a representative use case for a better privacy implementation is represented in the Windows 10 per-network randomization design framework. While this capability is also available in Android 9 it requires an advanced user to enable developer options and follow-on configuration settings, inevitably this is uncommon. Similarly, while the use of other temporary identifiers such as the \ac{TMSI} can be used by \ac{GSM} to afford a per-connection obfuscation, the \ac{IMEI} can still be obtained and tracked by the observer with low-cost commodity off the shelf systems~\citep{van2015defeating, mjolsnes2017easy, paget2010practical, strobel2007imsi}.

A second information leak, reveals the mobile device type, again this is due to the nature of the structured and regulated allocation of layer-2 hardware identifiers. The \ac{IMEI} in which the first 8 digits represents a \ac{GSMA} allocated \ac{TAC} maps directly to the exact manufacturer and model of the device. A \ac{MAC} address, has a three-byte prefix allocated by the \ac{IEEE} called the \ac{OUI} which indicates the manufacturer of the device.

\subsection{Definitions}
\label{sec:intro-defs}

\todo{Do we need the term ``signal''?}

In this paper, we consider mobile \emph{devices} such as phones, tablets, and laptop computers equipped with wireless communication technologies that enable each device to emit \emph{signals} consisting of data \emph{frames} following one or more \emph{protocols}, which determine characteristics of the signals and frames (such as signal range, frame rate, and frame content).

An \emph{event} refers to the observation of a frame emitted from a particular device, following a particular protocol, from a particular location, and at a particular time. For the purposes of this work, the ID, extracted from the frame, is assumed to be unique to the device and protocol via which it was sent.
A location may be represented as a single geospatial point, a geospatial region, or a geospatial probability distribution.

A \emph{trajectory} is a time-ordered sequence of events with the same ID (i.e. corresponding to the same device and protocol).

A \emph{trace} is the ``ground truth'' continuous spatio-temporal path followed by a device. Under ideal conditions, the events in a trajectory would correspond exactly to points along a trace, but in practice this is often not the case.



\todo{Two signals are \emph{co-traveling} at a given time if their corresponding devices are in high spatial proximity.}

\subsection{Problem Statement}
\label{sec:intro-problem}

\textbf{Objective}: Given a set of events corresponding to observations of signals emitted by mobile devices across different protocols, identify pairs of IDs corresponding to the same device.

We focus on the following two problems:
\begin{itemize}
	\item \textbf{Specific Query:} Given the ID of a target signal, return a ranked list of IDs likely corresponding to the same device as the target ID.
	\item \textbf{General Query:} Return a ranked list of pairs of IDs likely corresponding to the same device.
\end{itemize}


In the following section, we present a computational framework called SEXTANT for addressing such queries.

\subsection{Contributions and Outline}
\label{sec:intro-contrib}

The contributions of this paper can be summarized as follows:
\begin{itemize}
  \item A computational framework that couples a distributed architecture with novel time and space-efficient algorithms to perform spatio-temporal correlation of trajectories at scale and identify likely pairs of interest.
  \item Three spatio-temporal correlation measures designed to capture common pattern-of-life behavior: co-traveling likelihood, temporal coverage, and spatial coverage. Each is based solely on the observed time-stamped locations of entities, satisfies desired mathematical properties, and smoothly handles uncertainty in the data.
  \item A protocol-based pruning and scoring technique that dramatically improves both computational efficiency (in runtime and candidates considered) and accuracy (in precision and recall).
  \item A simulation model for generating spatio-temporal trajectories representing human mobility and device communication/signaling behavior.
  \item An application of the algorithms to a city-scale entity resolution and tracking problem, with results that highlight the continued need for robust anonymization techniques of network protocol hardware identifiers.
\end{itemize}

Section~\ref{sec:related} provides a survey of related literature. Section~\ref{sec:methods} introduces the SEXTANT framework, including computational approach and algorithms. Section~\ref{sec:eval} presents an experimental methodology and results. Section~\ref{sec:concl} concludes with discussion of results and future work.

%% file: 20-related.tex
\section{Related Work}
\label{sec:related}

\subsection{Multi-Protocol Entity Resolution}
\label{sec:related-cyber}

Existing privacy research on correlating entities based on signaling behavior spans several research areas. Some methods aim to recognize multiple device identifiers as belonging to a single device by linking across network protocols, gleaning details retrieved from upper layer network protocol information~\citep{martin2013correlating, longo2018pairing, rajavelsamy2018privacy, hong2018guti, o2017mobile}, as well as using signaling-based spatial-temporal attributes~\citep{martin2013correlating}. We describe in more detail the general body of work regarding spatio-temporal modeling in Section~\ref{sec:related-spatiotemporal}.

\todo{Don't say how we use the techniques here; just describe what they are.}

Extending on these works we rely heavily on the research of~\citet{martin2016decomposition} in order to provide granular model details for 802.11 \ac{MAC} addresses. This prior work, in which the authors utilize data extracted from higher layer protocols to build a graph of the allocation space for device manufacturers allows for fine-grained device type inference. We build a data corpus using this methodology for determining and pruning candidate pairs, described in detail in Section~\ref{sec:methods-device}.

An additional body of work, where the inclusion and ordering of optional management frame parameter fields / attributes can be utilized to construct a semi-unique \emph{device signature}, illustrates another method to identifying granular device model information~\citep{gentry2016passive}. Furthermore, this technique was utilized to aide in defeating randomization of \ac{MAC} addresses~\citep{martin2017study}. We construct a more exhaustive corpus of device signatures using the same methods as described by~\citet{gentry2016passive} and utilize the results for our candidate pair algorithm detections described in Section~\ref{sec:methods-device}.

\subsection{Spatio-Temporal Modeling and Correlation}
\label{sec:related-spatiotemporal}

\todo{Bulk this section up significantly.}

Barab{\'a}si and collaborators have used cell phone mobility data, at cell tower granularity, to study human mobility patterns. Gonz{\'a}lez et al.\ demonstrate the existence of scaling laws and other common properties of human mobility~\citep{gonzalez2008understanding}. Song et al.\ extend that work to examine parameters characterizing and distinguishing between the behavior of different individuals~\citep{song2010modelling}. Song et al.\ examine the predictability of an individual's location based on their past history, also proposing ways to extrapolate from aggregate data to accommodate sparse or missing information~\citep{song2010limits}. Wang et al.\ consider the task of predicting the formation of new social links using measures of ``mobile homophily,'' which quantify the similarity between individuals' location profiles~\citep{wang2011human}.

Jurdak et al.\ use higher-granularity locational data from Twitter geotags to validate hypotheses about human mobility and characterize individuals' movement patterns~\citep{jurdak2015understanding}.

Backes et al.\ propose an algorithm to infer which users of a social network are friends based on the spatial distributions of users' check-in locations~\citep{backes2017walk}. Thompson et al.\ infer pairwise influence between entities by relying only on the times of their observed individual activity~\citep{thompson2015inferring}.



However, although there is significant literature on studying aggregate patterns of human mobility, modeling and characterizing individual mobility patterns, correlating entities based on their spatial footprint, and time series correlation, few approaches have been proposed that utilize both spatial and temporal information to identify pairs of entities with high spatial proximity over time.

Furthermore, such spatio-temporal correlation methods are typically disjoint from methods for multi-protocol entity resolution using device identification techniques. In this work, we present algorithms that fundamentally integrate the spatial, temporal, and network protocol domains, along with a computational framework to perform such algorithms efficiently and at scale.

%% file: 30-methods.tex
\todo{is table of notation necessary? does it need to be updated/revised?}

\begin{table}[!ht]
	\caption{Notation}
	\label{tab:notation}
	\centering
	\begin{tabular}{c|l}
		\toprule
		Variable & Description \\
		\midrule
		$I$ & set of unique IDs, $\id \in I$ \\
		$E$ & set of events in the dataset, $e = (\id, \sigma, \tau) \in E$ \\
		$E_\id$ & set of events with ID $\id$, AKA the trajectory of ID $\id$ \\
		$S$ & set of spatial regions, $s \in S$ \\
		$T$ & set of time blocks/intervals, $t \in T$ \\
		$S \times T$ & set of spatio-temporal (S-T) cells, $(s, t) \in S \times T$ \\
		$E_{(s, t)}$ & set of events in S-T cell $(s, t)$ \\
		$E_{\id, (s, t)}$ & set of events with ID $\id$ in S-T cell $(s, t)$ \\
		$n$ & number of trajectories/IDs in the dataset, $n = |I|$ \\
		\bottomrule
	\end{tabular}
\end{table}

\section{Methods}
\label{sec:methods}

We now present SEXTANT (Scalable and Efficient Correlation for Spatio-Temporal Trajectory ANalyTics), a computational framework for
studying correlations in spatio-temporal behavior. Notation is provided in Table~\ref{tab:notation}.

\subsection{Approach}
\label{sec:methods-approach}

The SEXTANT workflow can be summarized by the following steps:
\begin{enumerate}
	\item Project events onto a discretized space and time.
	\item Determine a set of candidate ID pairs.
	\item Prune on device information embedded in frames.
	\item Compute correlation measures on remaining pairs.
	\item Rank the pairs based on their correlation scores.
\end{enumerate}
The algorithms are designed so that all computations can be performed efficiently over a distributed architecture in either an online/streaming or offline/batch setting. The steps are described in greater detail below.

\subsubsection{Discrete Spatio-Temporal Projection}
\label{sec:methods-approach-projection}




\begin{figure*}[t]
	\centering
	\subfloat {
		\label{fig:trajectories}
		\includegraphics[width=0.45\textwidth]{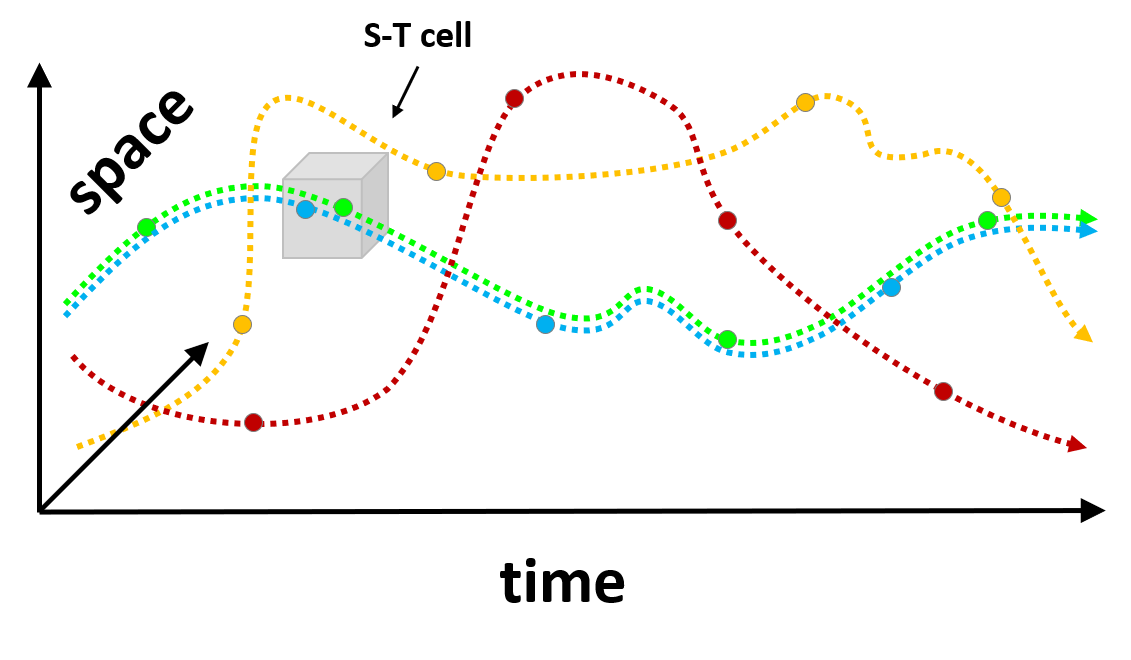}}
	\hfill
	\subfloat {
		\label{fig:st-volumes}
		\includegraphics[width=0.5\textwidth]{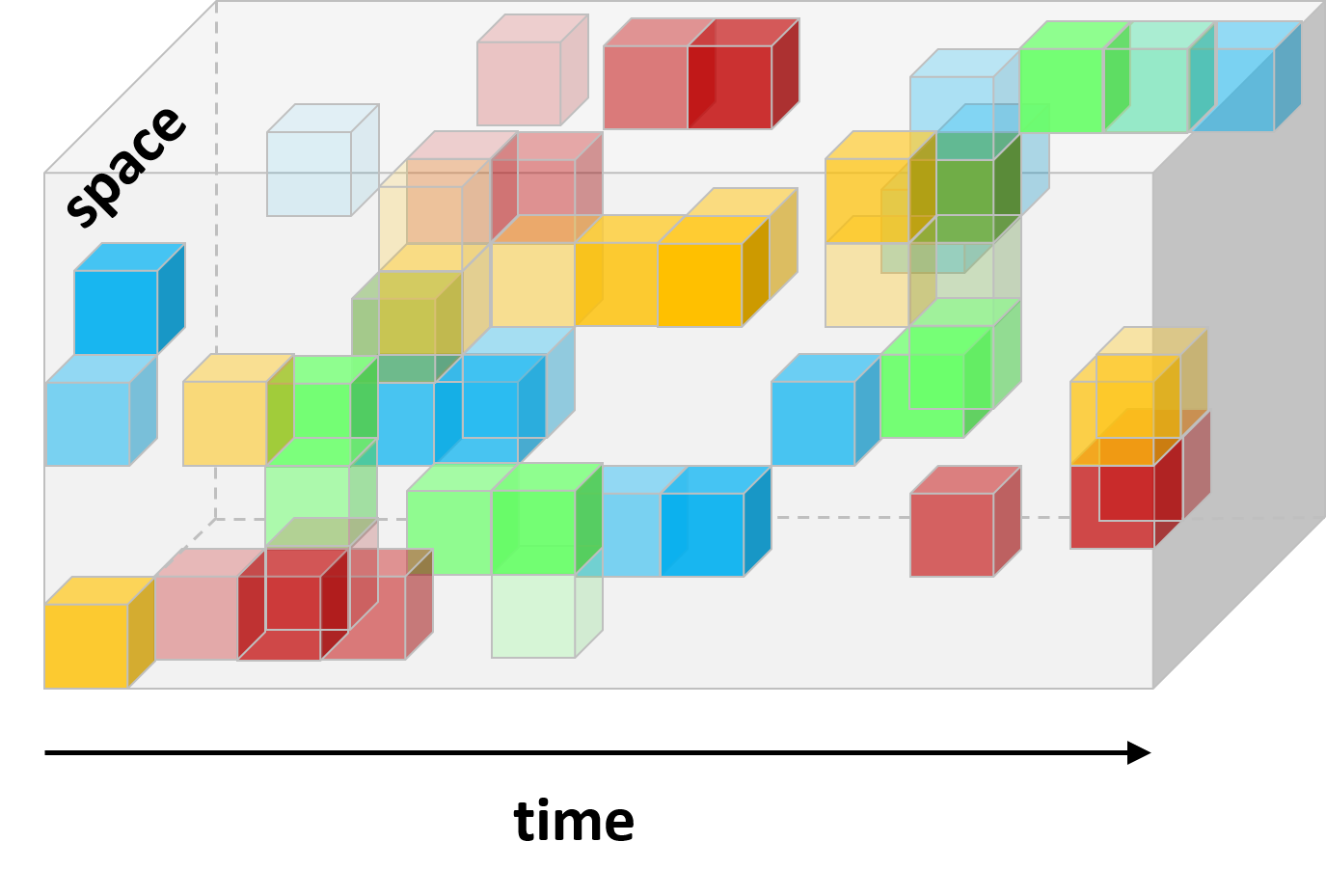}}
	\caption{(a) Four trajectories, two of which (blue and green) correspond to co-traveling sources. (b) Four S-T volumes corresponding to four trajectories. Shading indicates the weight of each S-T cell.}
	\label{fig:st-volume}
\end{figure*}

The first step of SEXTANT is to project events from a continuous to a discretized space and time, which allows for more efficient computation of overlap and correlation of trajectories. The surface of the Earth and the duration of the dataset are partitioned into spatio-temporal cells, or \emph{S-T cells}, each representing a specific spatial region on Earth during a specific time interval. Any spatio-temporal point can then be projected onto its corresponding S-T cell. For the spatial projection, we use the S2 Geometry library~\citep{s2geometry}, which maps locations on Earth to two-dimensional spatial regions whose boundaries are formed by geodesic curves, with up to centimeter-scale granularity. For the temporal projection, we partition time into fixed-length intervals. Spatial and temporal resolution and other parameters can be specified based on context and application constraints.

As described in Section~\ref{sec:intro-defs}, each event is specified by an (ID, time, location) tuple, which can be mapped to a weighted set of one or more S-T cells corresponding to its time and location, which we refer to as a (weighted) spatio-temporal volume, or \emph{S-T volume}. The weight of an S-T cell within an S-T volume represents the likelihood that the source was in the corresponding spatial region during the corresponding time interval.

A trajectory can also be modeled as an S-T volume. Since a trajectory is a sequence of events with the same source, the S-T volume for a trajectory results from taking the union of the S-T volumes for the corresponding events. Note that multiple events in a trajectory could have non-zero weights for the same S-T cell. Because the weight of an S-T cell represents the likelihood that the source was in the corresponding spatial region during the corresponding time interval, and distinct events are assumed to represent independent observations of the source's time and location,\footnote{We treat uncertainties in event locations as being independent. However, we do not assume that the event locations themselves are independent.} the combined weight of an S-T cell $(s,t)$ resulting from the union of multiple S-T volumes $V_1, \ldots, V_k$ is computed as $1 - \prod_{1 \leq i \leq k} \(1 - V_i[(s,t)]\)$, where $V[(s,t)]$ is the weight of S-T cell $(s,t)$ in S-T volume $V$. Examples of S-T volumes representing trajectories are illustrated in Figure~\ref{fig:st-volume}.

\subsubsection{Determining Candidate Pairs}
\label{sec:methods-approach-candidates}

The second step of our approach is to select pairs of sources that are likely candidates for being co-travelers. The challenge is to find an appropriate balance between being too inclusive, unnecessarily increasing the computational time, and being too selective, possibly overlooking true co-traveling pairs. A baseline criterion could be only considering pairs whose S-T volumes have non-empty overlap, i.e. there is at least one S-T cell for which both sources have non-zero weight.

\subsubsection{Pruning on Device Information}
\label{sec:methods-approach-pruning}

The goal of the pruning step is to use device information embedded in signal data to eliminate candidate signal pairs that are unlikely to originate from the same device, thus limiting the number of times that the computationally expensive calculation of correlation measures needs to be performed. Device information can further be leveraged to provide additional context for candidate pairs that do not get pruned. Alternatively, cross-validating results using spatio-temporal and device-related information could help identify spoofed or falsified information. More details on this step are provided in Section~\ref{sec:methods-device}.

\subsubsection{Computing Spatio-Temporal Correlation}
\label{sec:methods-approach-correlation}

Once candidate pairs have been selected and pruned, the spatio-temporal correlation between each pair is computed. We propose three spatio-temporal correlation measures that collectively are designed to identify co-travelers exhibiting common pattern-of-life behavior:
\begin{itemize}
	\item Co-traveling likelihood --- what fraction of the time two sources are co-traveling
	\item Temporal coverage --- measures the temporal span and frequency of their overlap
	\item Spatial coverage --- measures the spatial span and density of their overlap
\end{itemize}
The measures are defined and explained in Section~\ref{sec:methods-measures}.

\subsubsection{Ranking Candidate Pairs}
\label{sec:methods-approach-ranking}

After computing the three correlation measures on each candidate pair, the last step is to rank the candidate pairs. Often resources are limited, and ranking helps to prioritize potential co-travelers for further scrutiny. Although candidate pairs could be ranked by any one of the correlation measures individually and filter based on one or more of the individual scores, the best candidates are those who simultaneously have high co-traveling likelihood, temporal coverage, and spatial coverage. We aim to capture this in a single combined score.

The first step in generating the combined score is to scale all measures to the same range. The co-traveling likelihood is always in the range $[0, 1]$, whereas the temporal coverage and spatial coverage are in the range $[0, \infty)$. Possible mapping functions from $[0, \infty)$ to $[0, 1]$ that preserve ordering include $f(x) = \min(\frac{x}{\alpha}, 1)$ for $\alpha > 0$ and $f(x) = 1 - \frac{1}{\beta^x}$ for $\beta > 1$.

The second step is to combine the scaled values. Possible combining functions include the minimum, the arithmetic mean, and the geometric mean. Either the arithmetic or geometric mean could be modified by assigning weights to the correlation measures to reflect desired priorities.
\subsection{Comparing Device Information}
\label{sec:methods-device}

We present procedures for both pruning and scoring candidate ID pairs based on the compatibility between information leaked from two common wireless communication protocols: GSM and WiFi/802.11. Although we focus on GSM and WiFi here, similar techniques may apply to other protocols as well.

\subsubsection{Pruning Incompatible Pairs}
\label{sec:methods-device-pruning}

As described in Section~\ref{sec:intro-background}, both GSM and WiFi frames contain device-related information. Both the device manufacturer and model can be determined from the TAC portion of the IMEI in GSM frames. Similarly, the OUI portion of the MAC address in WiFi frames maps to the device manufacturer. WiFi frames may also contain WPS fields indicating the manufacturer and model of the device.

Given the set of candidate ID pairs, SEXTANT uses the device information extracted from GSM and WiFi frames to eliminate incompatible pairs. If the manufacturers corresponding to the two identifiers do not match, or if WiFi frames with the WPS model field are observed and the models do not match, then those IDs are said to be incompatible, and the ID pair can be pruned from the candidate set.

\subsubsection{Device Match Scores}
\label{sec:methods-device-scores}

When it is not explicitly represented in the frames of the observed signals, SEXTANT leverages two existing techniques to infer manufacturer and model information.

The first technique is based on a highly specific device signature derived from a subset of the parameters present in 802.11 management frames~\citep{gentry2016passive}. The approach entails populating a table of known signatures and their manufacturer and model information. When a new identifier is observed, the manufacturer and model information for other identifiers with the same signature is used to infer a probability distribution over possible manufacturers and models for the device corresponding to the new identifier.

The second technique also uses a lookup table but populates it with MAC addresses instead of signatures~\citep{martin2016decomposition}. It also differs from the first technique because, whereas many identifiers may have the same signature, MAC addresses are assumed to be unique. To compensate, it exploits inherent structure in how MAC addresses are assigned. For example, in addition to the first three bytes (the OUI) being specific to the manufacturer, manufacturers often allocate consecutive blocks of MAC addresses to devices of the same model. When a new MAC address is observed, the manufacturer and model information for nearby entries in MAC address space is used to infer a probability distribution over possible manufacturers and models for the device corresponding to the new MAC address.

SEXTANT uses these probability distributions to assign a device match score to each candidate ID pair. If WiFi probe requests with the model WPS field are observed and there is a match for both manufacturer and model, the ID pair is assigned a score of 1. Otherwise, SEXTANT assigns the score to be a function (e.g. the maximum, or the average) of the match probabilities based on the signature and MAC address inference techniques.

Because the inference methods rely on incomplete information, collected from samples of the large and expanding population of device manufacturers and models, SEXTANT does not consider a high score to be definitive proof of a device match, nor does it consider a low score to be sufficient justification for pruning a candidate pair. However, these methods can still provide useful context about the likelihood that two signals are coming from devices with the same manufacturer and model. For example, it can be a good differentiator in prioritizing between several candidate ID pairs with similar spatio-temporal correlation scores, or it can increase confidence that two IDs with a high correlation score are indeed coming from the same device.

Regardless of the techniques used, leaked information about device properties can be used in conjunction with spatio-temporal correlation measures to enhance SEXTANT's performance and utility.

\subsection{Spatio-Temporal Correlation Measures}
\label{sec:methods-measures}

At the core of SEXTANT is the ability to quantify spatio-temporal correlation between two trajectories, which is achieved by comparing the corresponding S-T volumes. As stated in Section~\ref{sec:intro-problem}, the goal of this work is to identify co-traveling sources exhibiting common pattern-of-life behavior --- that is, sources whose traces often have high spatial proximity, across a diversity of space and time. In terms of S-T volumes, this means having overlapping S-T cells with high weight, corresponding to a high likelihood that both sources were in approximately the same place at approximately the same time.

We propose three spatio-temporal correlation measures that collectively are designed to identify such pairs: co-traveling likelihood, temporal coverage, and spatial coverage. Depending on the context, a user of SEXTANT may choose to focus on only a subset of these measures or to consider additional measures.

\subsubsection{Co-Traveling Likelihood}
\label{sec:methods-measures-likelihood}

The \emph{co-traveling likelihood} aims to measure the fraction of time during which two sources are co-traveling.

First consider an ideal world in which the exact and complete traces of entities are known, so that there is an event $(i_a, \sigma, \tau)$ indicating the exact location $\sigma$ of each source $a$ at every moment in time $\tau$. Then we could measure the fraction of time during which two sources $a$ and $b$ were exact co-travelers (i.e.\ in the exact same place at the exact same time). We call this measure the \emph{exact co-traveling likelihood} because it represents the likelihood that the two sources were co-traveling at a point in time selected uniformly at random between $\tau_\text{min}$ and $\tau_\text{max}$.

In many real-world applications, it may be desirable to consider a definition of co-travelers that allows sources to be nearby rather than exactly co-located. To accommodate this, we can apply a discrete spatio-temporal projection as described in Section~\ref{sec:methods-approach-projection}, and then measure the fraction of time intervals during which two sources $a$ and $b$ were located in at least one common spatial region. We call this measure the \emph{discrete co-traveling likelihood}.

Next, consider a less ideal world in which events still have perfect accuracy and precision -- that is, each event corresponds to the correct location of an entity at a given time -- but there are only a finite number of events per source rather than a complete trace. Then the above formulation would be strongly biased against sources with infrequent events, as even two perfectly co-traveling sources may only occasionally be observed during the same time interval. This bias can be addressed by conditioning on the existence of observations of $a$ and $b$, yielding a measure which we call the \emph{conditional discrete co-traveling likelihood}.

Finally, consider a more realistic world in which, additionally, event locations may be inexact, resulting in S-T volumes composed of S-T cells with non-binary weights. As described in Section~\ref{sec:methods-approach-projection}, the weight of a cell in an S-T volume corresponds to the estimated likelihood that a particular source was in a particular spatial region during a particular time interval. In adapting the above formulation, this uncertainty in event locations affects our representations of both the target and conditioning variables. We call the resulting measure the \emph{estimated conditional discrete co-traveling likelihood}.

Throughout the paper, we use the term \emph{co-traveling likelihood} ($CTL$) to refer to the estimated conditional discrete co-traveling likelihood. The co-traveling likelihood between two sources $a$ and $b$ can be expressed as follows:
\begin{equation}
	\label{eqn:ctl-theory}
	CTL(a,b) = \frac{\sum_{t \in T} P({AB}_t)}{\sum_{t \in T} P(A_t, B_t)},
\end{equation}
where ${AB}_t$ denotes the statement that there is at least one spatial region $s$ in which both sources $a$ and $b$ were observed during time interval $t$, and $A_t$ (resp.\ $B_t$) denotes the statement that source $a$ (resp.\ $b$) was observed at least once during time interval $t$. As stated above, we treat all uncertainties in event locations as being independent (but we do not assume that the event locations themselves are independent). Under that assumption, we have the following:
\begin{gather}
	P(AB_t) = 1 - \prod_{s \in S} (1 - P(A_{s,t}) \cdot P(B_{s,t})) \\
	P(A_t, B_t) = P(A_t) \cdot P(B_t) \\
	P(A_t) = 1 - \prod_{s \in S} (1 - P(A_{s,t})) \\
	P(B_t) = 1 - \prod_{s \in S} (1 - P(B_{s,t})),
\end{gather}
where $A_{s,t}$ (resp.\ $B_{s,t}$) denotes the statement that source $a$ (resp.\ $b$) was observed to be in spatial region $s$ at least once during time interval $t$. Following the discrete projection approach detailed in Section~\ref{sec:methods-approach-projection}, we represent the probabilities $P(A_{s,t})$ and $P(B_{s,t})$ using the weights in the corresponding S-T volumes, i.e. $P(A_{s,t}) = V_a[(s,t)]$ and $P(B_{s,t}) = V_b[(s,t)]$. Substituting back into Equation~(\ref{eqn:ctl-theory}) yields the following formula for computing the co-traveling likelihood
$CTL(a,b)$
between two sources $a$ and $b$, given their corresponding S-T volumes:
\begin{equation}
	\label{eqn:ctl}
	\frac{\displaystyle\sum_{t \in T} \: \(1 - \displaystyle\prod_{s \in S} (1 - V_a[(s,t)] V_b[(s,t)])\)}{\displaystyle\sum_{t \in T} \: \(1 - \displaystyle\prod_{s \in S} (1 - V_a[(s,t)])\) \(1 - \displaystyle\prod_{s \in S} (1 - V_b[(s,t)])\)}.
\end{equation}


\subsubsection{Temporal Coverage}
\label{sec:methods-measures-temporal}

\begin{figure}[t]
	\centering
	\includegraphics[width=0.35\textwidth]{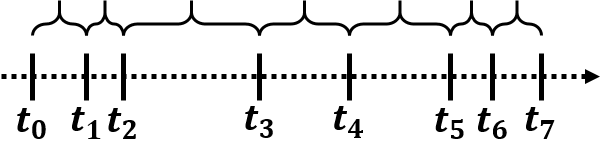}
	\caption{Temporal coverage is computed based on the differences in time between consecutive co-occurrences.}
	\label{fig:temporal-coverage}
\end{figure}

To identify sources with common pattern-of-life behavior, we want to reward pairs of trajectories whose overlap both spans a significant period of time and has sufficiently frequent co-occurrences (when both trajectories have events in the same location at the same time) throughout the span. The following properties attempt to capture this intuition:
\begin{enumerate}[T1:]
	\item Given a sequence of co-occurrence times, adding more times to the end will increase the score.
	\item Given a sequence of co-occurrence times, adding more times in the middle will increase the score.
	\item Given a sequence of co-occurrence times, scaling up all times will increase the score.
	\item Given a time span, the maximum score achievable for a sequence of co-occurrence times lying entirely within the span is finite.
\end{enumerate}

We use the differences in time between consecutive co-occurrences as the basic building blocks (see Figure~\ref{fig:temporal-coverage}). Let $t_0, \ldots, t_k$ be the times of event co-occurrences. Then the temporal coverage is defined as follows:
\begin{equation}
	\mathtt{temporal\_coverage}(t_0, \ldots, t_k) = \sum_{i=1}^k \log((t_i - t_{i-1}) + 1),
\end{equation}
where differences in time are measured in units of days. This satisfies all four desired properties.

One simple and efficient way to accommodate uncertainty is to only count times at which the conditional likelihood of co-occurrence (i.e. $P({AB}_t) / P(A_t, B_t)$) exceeds a fixed threshold. Alternatively, one could compute a weighted sum that uses the Chain Rule to probabilistically consider each possible sequence of true co-occurrences.

Note that although for the purposes of this paper we compute the temporal coverage on a sequence of co-occurrences, it could equivalently be applied to any sequence of timestamps.

\subsubsection{Spatial Coverage}
\label{sec:methods-measures-spatial}

\begin{figure}[!htb]
	\centering
	\includegraphics[width=0.3\textwidth]{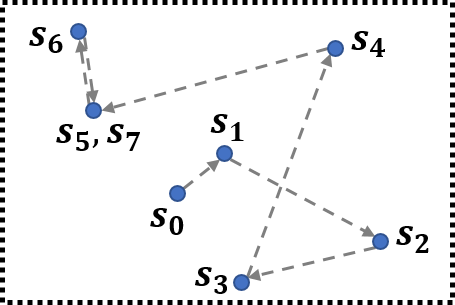}
	\caption{Spatial coverage is computed based on the distances between consecutive co-occurrences.}
	\label{fig:spatial-coverage}
\end{figure}

To identify sources with common pattern-of-life behavior, we also want to reward pairs of trajectories whose overlap both spans a significant spatial region and
demonstrates sufficiently long distances traversed within the region. The following properties attempt to capture this intuition:
\indent
\begin{enumerate}[S1:]
	\item Given a sequence of co-occurrence locations, adding more locations to the end will increase the score.
	\item Given a sequence of co-occurrence locations, adding more locations in the middle will increase the score.
	\item Given a sequence of co-occurrence locations, scaling up the spatial coordinate values for all locations will increase the score.
	\item Given a spatial region, the maximum score achievable for a sequence of co-occurrence locations lying entirely within the region is finite.
\end{enumerate}
There is a natural extension of the formula for temporal coverage to the spatial domain, substituting the distances between consecutive co-occurrences for the elapsed times between them. However, due to the higher dimensionality and degrees of freedom of an ordered sequence of spatial locations as opposed to an ordered sequence of times, the corresponding version of the formula for temporal coverage does not work for spatial coverage: it satisfies Properties S1, S2, and S3, but not S4. For example, a pair of co-traveling sources can travel between the same two locations an arbitrary large number of times, resulting in an arbitrarily high score, even if the two locations are very close together. Furthermore, it would unduly favor a single one-time path for a pair of very high-frequency sources (e.g. two people sitting next to one another on a train) over several co-occurrences spread out over space and time for a pair of low-frequency sources (which more likely represents pattern-of-life behavior).


Two modifications to the formula address these problems:
\begin{enumerate}
	\item We impose an upper bound based on how much coverage of a region is sufficient to demonstrate pattern-of-life behavior; that is, after consistent and repeated co-traveling behavior has been established over a region, the information gained from additional co-traveling within that region does not provide much marginal value. For example, after a pair of sources has jointly traversed the equivalent of several times the length of the perimeter of a spatial region while remaining inside that region, it may be reasonable to consider pattern of life to have been established. (Considering the area of the region rather than its perimeter would unduly penalize sets of spatial locations consisting of two tight clusters of points, even if they were far apart).
	\item We aggregate the distances between consecutive co-occurrences using a simple sum, rather than the sum of the logs of the distances, applying a composition of monotonic functions (thus preserving order) only at the end to maintain desired scaling properties.
\end{enumerate}

We use distances between consecutive co-occurrences as the basic building blocks (see Figure~\ref{fig:spatial-coverage}). Let $s_0, \ldots, s_k$ be the locations of event co-occurrences. Then the spatial coverage is defined as follows:
\begin{equation}
	\mathtt{spatial\_coverage}(s_0, \ldots, s_k) = \ln\(\min\(\sum_{i=1}^k \Delta s_i, \; \gamma \cdot Perim(\{s_0, \ldots, s_k\})\) + 1\),
\end{equation}
where $\Delta s_i = distance(s_{i-1}, s_i)$, $\gamma$ is a pre-determined constant,
$Perim()$ denotes the perimeter of the convex hull of a set of points,
and distances are measured in units of kilometers. This satisfies all four desired properties. Property S1 is strongly satisfied (i.e. the score strictly increases when additional locations are added) if the sum is less than the perimeter of the convex hull, or if the additional location is outside the previous convex hull, and is weakly satisfied (i.e. the score remains the same, rather than increasing or decreasing) otherwise. Property S2 is strongly satisfied if the sum is less than the perimeter of the convex hull and the additional location is not co-linear with its chronologically preceding and succeeding locations, or if the additional location is outside the previous convex hull, and is weakly satisfied otherwise. Properties S3 and S4 are always strongly satisfied.

Uncertainty over whether there was at least one co-occurrence during a given time interval can be handled as described above for temporal coverage. Multiplicity of co-occurrence locations within a time interval can be addressed by finding the centroid, or average location, of the co-occurrence locations. Uncertainty within a time interval $t$ can be accommodated by taking a weighted average location as follows:
\begin{equation}
	\<\frac{\sum_s P(A_{s,t}) \cdot P(B_{s,t}) \cdot x_s}{\sum_s P(A_{s,t}) \cdot P(B_{s,t})}, \frac{\sum_s P(A_{s,t}) \cdot P(B_{s,t}) \cdot y_s}{\sum_s P(A_{s,t}) \cdot P(B_{s,t})}\>,
\end{equation}
where $\<x_s, y_s\>$ is the center of spatial cell $s$.

To improve efficiency, approximation algorithms can be used to estimate the perimeter of the convex hull. For example, maintaining a bounding rectangle around the co-occurrence locations over-estimates the perimeter of the true convex hull by at most a factor of $\sqrt{2}$. See Appendix~\ref{apx:proof} for proof.

Note that although for the purposes of this paper we compute the spatial coverage on a sequence of co-occurrences, it could equivalently be applied to any sequence of locations, for example to measure the spatial coverage of an individual source.

\subsection{Algorithm}
\label{sec:methods-algs}

%
%
%

The SEXTANT algorithm includes the following components:
\begin{enumerate}[(1)]
	\item For each candidate pair $i, i' \in I$, maintain the following values:
	\begin{itemize}
		\item The time interval and location of their last co-occurrence;
		\item The numerator for the co-traveling likelihood (the sum over the probability of co-occurrence for each time interval);
		\item The denominator for the co-traveling likelihood (the sum over the probability that both sources were observed for each time interval);
		\item The sum of the logs of the co-occurrence inter-event times;
		\item The sum of the co-occurrence inter-event distances; and
		\item The latitudes and longitudes determining the bounding box for the convex hull approximation.
	\end{itemize}
	\item For each new event $e = (\id, \sigma, \tau) \in E$, perform the following:
	\begin{itemize}
		\item Map $e$ to the corresponding S-T volume consisting of weighted S-T cells, optionally capturing any uncertainty in its observed location.
	\end{itemize}
	\item At the end of each time interval $t \in T$, perform the following:
	\begin{itemize}
		\item Identify new candidate pairs by comparing events that are mapped to the same S-T cell;
		\item Compare device information for new candidate pairs and prune if they conflict; and
		\item Initialize or update the values in Step (1) appropriately.
	\end{itemize}
	\item When results are desired:
	\begin{itemize}
		\item Compute the three spatio-temporal correlation measures (co-traveling likelihood, temporal coverage, and spatial coverage) using the values in Step (1);
		\item Combine the three correlation measures to yield a single correlation score; and
		\item Sort the correlation scores and return the results to the user along with relevant device information.
	\end{itemize}
\end{enumerate}

\subsection{Complexity Analysis}
\label{sec:methods-complexity}

%
%


The SEXTANT algorithm requires only a constant amount of storage space per candidate pair (see Step (1)). Step (2) takes constant time per event. Step (3) takes time linear in the number of relevant events in that time interval for each candidate pair, which is amortized constant time per event. Step (4) can be computed in time linear in the number of candidate pairs using efficient numerical sorting algorithms. Thus in total the SEXTANT algorithm runs in $O(\mathcal{C})$ space and $O(\mathcal{C} + |E|)$ time, where $\mathcal{C}$ is the number of candidate pairs and $|E|$ is the total number of events observed. All steps of the algorithm can be performed in parallel, utilizing the Map-Reduce paradigm to achieve scalability.

%
%
%
%
%
%
%
%
%
%
%
%
%
%
%
%
%
%
%
%
%

%% file: 41-evaluation.tex
\section{Evaluation}
\label{sec:eval}

To evaluate SEXTANT's performance, we develop a simulation model for generating spatio-temporal trajectories representing the combination of human mobility and wireless communication signals emitted by mobile devices.
We then perform experiments on the synthetic data generated by those models using an implementation of the SEXTANT framework and algorithms.

\subsection{Simulation Model and Synthetic Dataset}
\label{sec:eval-sim}

\begin{figure}[t]
	\centering
	\includegraphics[width=0.35\textwidth]{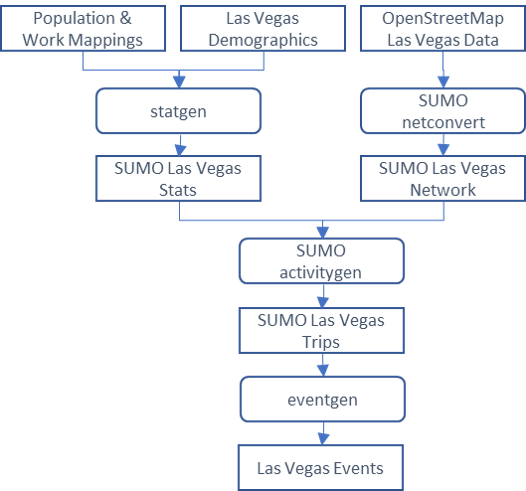}
	\caption{Process flow diagram for the Vegas Events simulation.}
	\label{fig:sumo-process}
\end{figure}

Although there are several publicly-available human mobility datasets, such as the GeoLife dataset~\citep{zheng2009mining},
\todo{and ?}
none of them have labeled pairs of trajectories coming from the same device or entity with which SEXTANT's performance could be evaluated. To that end, we develop a simulation model that integrates probabilistic models of human mobility, device properties, and signal and sensor characteristics to generate events.

We then use the simulator to generate a large set of events representing the activity of residents and visitors in the city of Las Vegas and their mobile devices, which we call the Vegas Events dataset.

The process is outlined in Figure~\ref{fig:sumo-process}. \todo{rework the flow diagram; include just SUMO piece or all?} Details on the different components of our simulation model are provided below.

\subsubsection{Human Mobility}

We use the open source Simulation of Urban MObility (SUMO) software, in particular the ActivityGen function, to simulate human mobility patterns~\cite{sumo}.
\todo{redo web citations}
Given a set of population parameters and a base map with infrastructure data, SUMO outputs sequences of start and end times and locations for trips that each person takes over the course of each day. In particular, each working adult is randomly assigned a home and work location sampled from spatial distributions based on road types (e.g. a home is more likely to be located along a residential street than along an arterial highway). In addition to home and work locations, which are visited regularly, each person may have one or more locations that are visited on an occasional basis (representing e.g. stores, restaurants, or movie theaters). Visitors (i.e. people who do not live or work in the geographical region being represented) are modeled as trips between randomly-selected locations.

\subsubsection{Device Properties}

Each working adult is assumed to carry a mobile device that emits both GSM and WiFi frames. To facilitate pruning on device information and the computation of device match scores, we generate synthetic device information using statistics from real-world data collections. Our simulator uses the following procedure to assign device-related properties to each person's mobile device:
\begin{enumerate}
	\item \label{step:man-model} Select a manufacturer and model
	\item \label{step:wps-indicator} Select whether WPS information is observed
	\item \label{step:oui-lookup} Select a proxy for MAC address
	\item \label{step:sig-lookup} Select a device signature
\end{enumerate}
We first calculate statistics based on real-world samples of WiFi frames collected from mobile devices to seed the randomized selection procedure. Two data collections were used for this purpose, based on the methods of \citet{martin2016decomposition} and \citet{gentry2016passive}, respectively --- the same collections used to form the lookup tables described in Section~\ref{sec:methods-approach-pruning}.

The first collection includes the following properties of each observed device: the manufacturer, the model, whether WPS information was observed, and the MAC address. The distribution of manufacturers and models is used for Step~\ref{step:man-model}. The total fraction of devices for which WPS information was observed is used for Step~\ref{step:wps-indicator}. The lookup table formed from the MAC addresses is used to estimate, for each manufacturer and model, a probability distribution over the manufacturer and model of the nearest neighbor in MAC address space with the same OUI, which is used for Step~\ref{step:oui-lookup}. Because SEXTANT only uses the MAC address for the nearest-neighbor lookup, rather than simulating an actual MAC address, we sample an inferred nearest-neighbor manufacturer and model from that distribution as a proxy. Ties for nearest neighbor are broken arbitrarily.

The second collection includes the following properties of each observed device: the manufacturer, the model, and a device signature. Step~\ref{step:sig-lookup} is performed by selecting a device signature uniformly at random from the set of signatures that were observed for devices of the selected manufacturer and model. If no signature was observed for a particular manufacturer and model, a unique signature is assigned that does not appear in the lookup table.

\subsubsection{Signal and Sensor Characteristics}
The final step in the simulation is to generate events representing observations of GSM and WiFi signals emitted by the devices. This process is dictated by parameters designed to capture three key aspects of signaling and sensing behavior: (a) the rate at which events are generated; (b) the spatial uncertainty in sensor observations; and (c) sensor coverage. Varying these characteristics across a synthetic dataset allows robustness to be evaluated, validating SEXTANT's usefulness for a multitude of real-world scenarios and use cases.

We model the rate at which events are generated by sampling inter-event times from a Bounded Pareto Distribution, a power-law distribution, with a device-specific shape parameter sampled from a protocol-specific distribution. For spatial uncertainty, we represent the observed location of each event as a two-dimensional Gaussian distribution of possible true locations, with major and minor axis lengths and orientation sampled from protocol-specific distributions, centered at a point sampled randomly from the distribution with the same shape but whose center is the true location of the simulated device at that time. Imperfect sensor coverage is modeled by randomly selecting a protocol-specific subset of geospatial regions within which no signals are observed, implemented by sampling from a Bernoulli random variable once for each geospatial region defined by the nearest OpenStreetMap road segment, whose parameter is specific to the road type (e.g. arterial highway or residential street).

\subsubsection{The Vegas Events Dataset}

Our simulation model generates a large synthetic dataset of more than 200 million events produced by 296,329 mobile devices carried by Las Vegas residents and visitors over a span of 30 days. SUMO/ActivityGen was seeded with OpenStreetMap infrastructure data~\cite{OpenStreetMap} and demographic data from the U.S. Census~\cite{censusLasVegas}. 

\begin{figure}[t]
	\centering
	\includegraphics[width=0.25\textwidth]{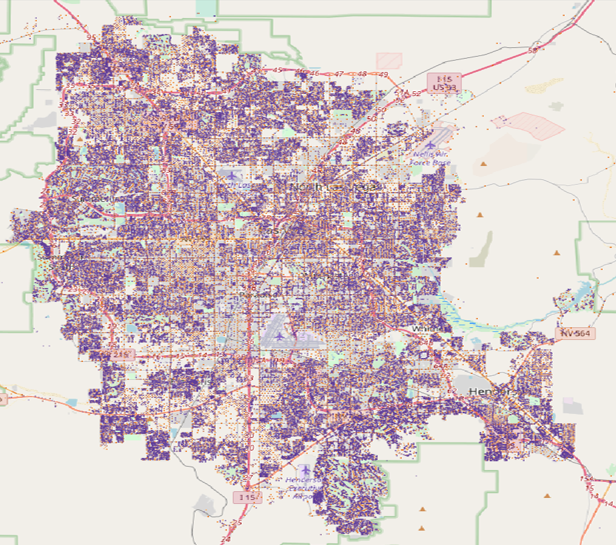}
	\caption{Spatial overview of the Vegas Events dataset.}
	\label{fig:spatial-overview}
\end{figure}
See Figure~\ref{fig:spatial-overview} for the spatial distribution of events.  Details about the parameters used in generating the Vegas Events dataset are provided in Appendix~\ref{apx:parameters}.

\subsection{SEXTANT Implementation}
\label{sec:eval-implementation}

For scalability, our experimental test bed was a Hadoop cluster running on 32 physical nodes, each allocated 40 vcores and 200 GB of memory. Algorithms were implemented in a combination of HiveQL and custom Java UDFs and were executed as Yarn jobs.


\todo{discretization: 3-sigma ellipses assuming bivariate normal/Gaussian distribution, 15-minute time intervals with 50\% discount factor}

For algorithm parameters, we used the threshold and spatial averaging techniques for handling uncertainty with a threshold of 0.5; the approximation algorithm for estimating the perimeter of the convex hull; a value of $\gamma = (e^5 - 1)/40$ in the definition of spatial coverage; and the following formula, the geometric mean over scaled versions of the three correlation measures, for computing the combined spatio-temporal correlation score:
\begin{equation}
	\sqrt[3]{CTL(a, b) \; \min\!\(\frac{TCov(a, b)}{10}, 1\) \; \min\!\(\frac{SCov(a, b)}{5}, 1\)}.
\end{equation}
When incorporating device match scores, we take the overall combined score to be the weighted average of the spatio-temporal correlation score and the maximum device match score over the two inference methods, with weights of 0.8 and 0.2, respectively.

\subsection{Experimental Setup}
\label{sec:eval-setup}

\subsubsection{Evaluative Metrics}

To evaluate performance for the specific query, we perform a query on every GSM identifier and calculate the fraction of IDs for which the true corresponding WiFi ID
has the top-ranked score.

To evaluate performance for the general query, we plot the precision-recall curve and compute the maximum F1-score, which help to visualize and quantify, respectively, how well a method is able to simultaneously achieve low rates of both Type I and Type II errors.

\subsubsection{Methods for Comparison}

\begin{figure*}[t]
	\centering
	\subfloat{\includegraphics[width=0.33\textwidth]{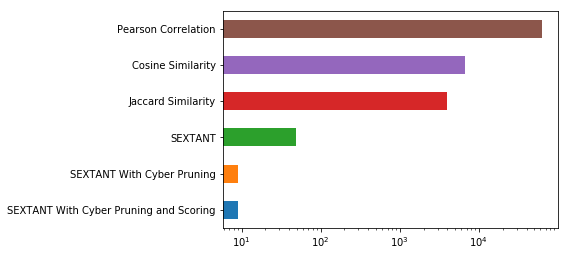}}
	\subfloat{\includegraphics[width=0.33\textwidth]{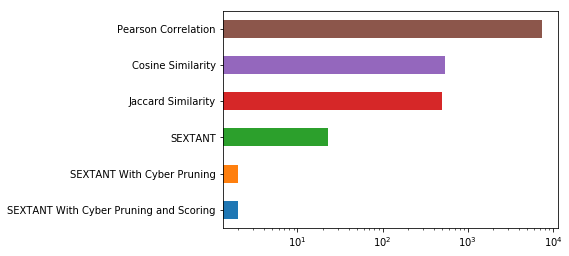}}
	\subfloat{\includegraphics[width=0.33\textwidth]{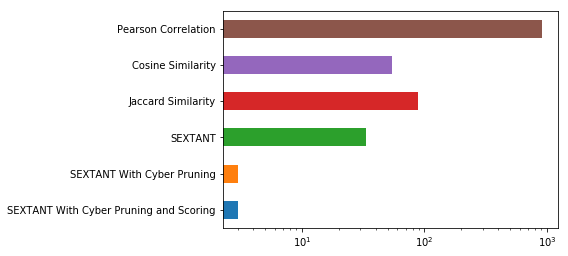}} \\
	\subfloat{\includegraphics[width=0.33\textwidth]{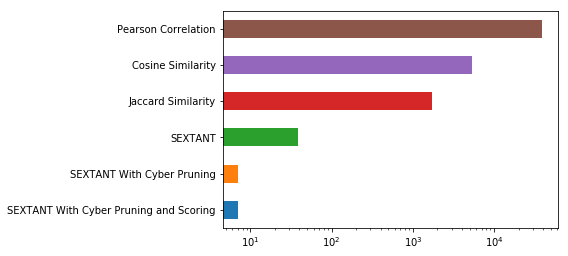}}
	\subfloat{\includegraphics[width=0.33\textwidth]{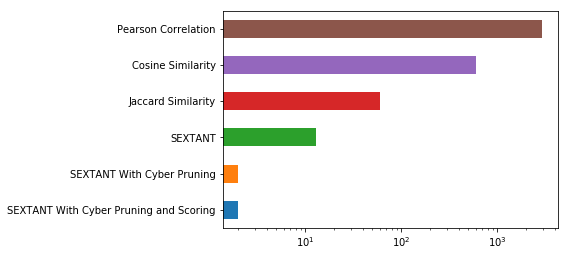}}
	\subfloat{\includegraphics[width=0.33\textwidth]{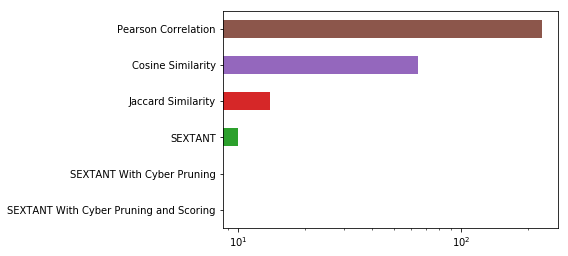}} \\
	\subfloat{\includegraphics[width=0.33\textwidth]{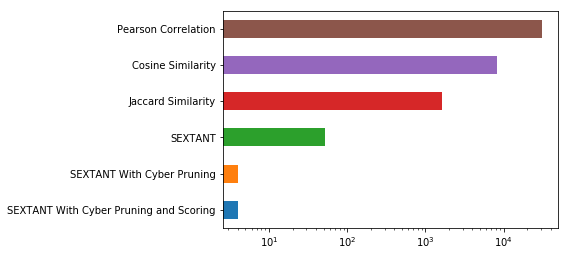}}
	\subfloat{\includegraphics[width=0.33\textwidth]{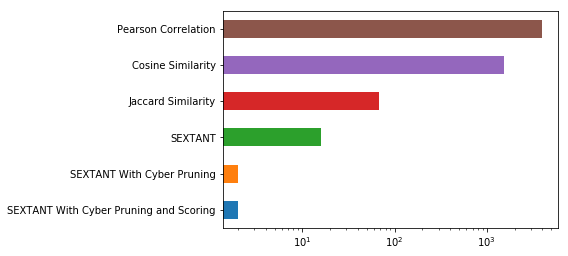}}
	\subfloat{\includegraphics[width=0.33\textwidth]{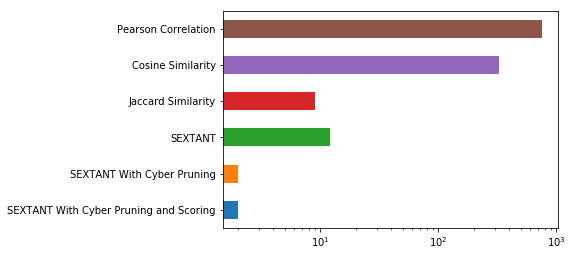}}
	\caption{Results for the specific query. Out of the 296,329 devices in the Vegas Events dataset, the number of GSM identifiers for which the WiFi identifier with the highest score did not correspond to the same device. Rows correspond to spatial granularity of levels 15, 16, and 17 in the S2 hierarchy, respectively. Columns correspond to temporal granularity of 5, 20, and 60 minutes. Results shown on a log scale.}
	\label{fig:specific-query-top1}
\end{figure*}

We compare SEXTANT's performance with that of several existing correlation methods: Pearson correlation, cosine similarity, and Jaccard similarity. For Pearson correlation and cosine similarity, the trajectories of the two IDs being compared are represented as real-valued vectors in a high-dimensional space, one dimension for each S-T cell. For Jaccard similarity, each trajectory is represented as the set of S-T cells with non-zero weight.

We also compare three different versions of the SEXTANT results: using the spatio-temporal algorithms alone without pruning, the spatio-temporal algorithms with pruning, and with the device match score incorporated into the combined score in addition to pruning.



\subsubsection{Spatial and Temporal Granularity}

To evaluate how sensitive the methods are to the granularity of the cells in the spatio-temporal projection, we perform experiments across a range of parameter values: spatial granularity at levels 15, 16, and 17 in the S2 hierarchy, corresponding to spatial cells of approximate side length 280 meters, 140 meters, and 70 meters, respectively; and time granularity of 5 minutes, 20 minutes, and one hour.

\subsection{Results}
\label{sec:eval-results}

Below we present the results from running SEXTANT on the Vegas Events dataset with varying parameters for spatial and temporal granularity, as well as experiments comparing SEXTANT to existing methods.

\subsubsection{Specific Query}
Figure~\ref{fig:specific-query-top1} shows the results from running a specific query for each GSM identifier in the Vegas Events dataset. For each method being compared, and for each choice of parameters for spatial and temporal granularity, the plot shows the number of GSM identifiers (out of the 296,329 devices in the Vegas Events dataset) for which the WiFi identifier with the highest score did not correspond to the same device.

The results indicate that SEXTANT with pruning based on device information significantly out-performs the comparison methods for all parameter choices. When using only spatio-temporal information, SEXTANT still out-performs the other methods in all cases except for the most granular setting, when the Jaccard similarity method performs slightly better. For all granularities, SEXTANT with pruning based on device information finds the correct GSM-WiFi pairs for all but at most 10 of the 296,329 devices.

\subsubsection{General Query}

\begin{figure}[!ht]
	\centering
	\includegraphics[width=.5\textwidth]{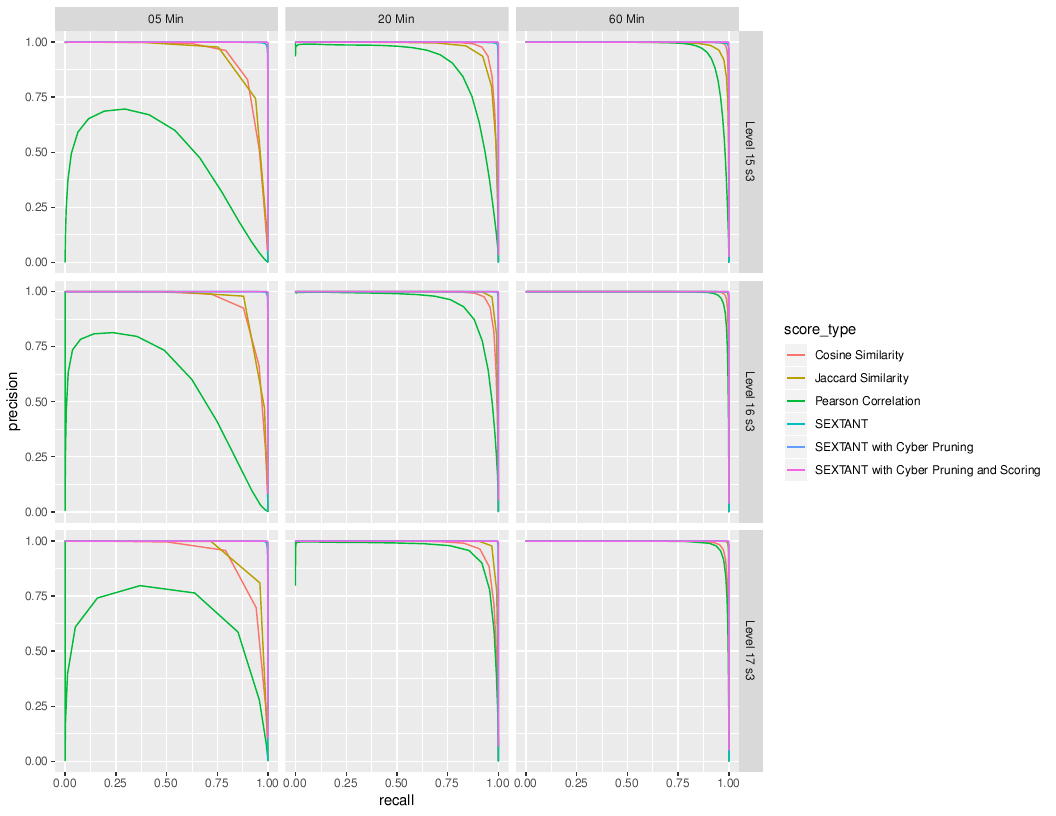}
	\caption{Precision-recall curves from running a general query on the Vegas Events dataset.}
	\label{fig:general-query-prcurves}
\end{figure}

\begin{figure*}[t]
	\centering
	\includegraphics[width=\textwidth]{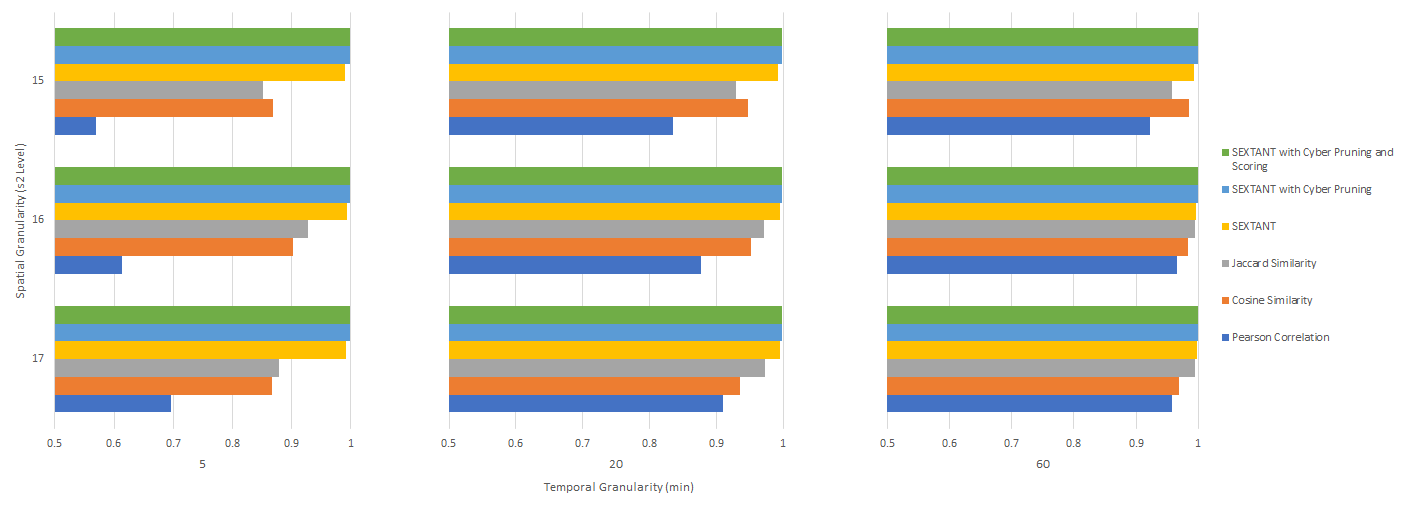}
	\caption{Maximum F1-scores achievable for the general query, by correlation method. Rows correspond to spatial granularity of levels 15, 16, and 17 in the S2 hierarchy, respectively. Columns correspond to temporal granularity of 5, 20, and 60 minutes.}
	\label{fig:f1}
\end{figure*}

Figures~\ref{fig:general-query-prcurves} and~~\ref{fig:f1} show the precision-recall curves and the maximum F1-scores, respectively, resulting from running a general query on the Vegas Events dataset using each correlation method. For each choice of parameters for spatial and temporal granularity, the precision-recall curve illustrates the tradeoff between Type I errors (false positives) and Type II errors (false negatives) in predicting whether each GSM-WiFi ID pair corresponds to the same device. The F1-score captures the best performance achievable with each correlation method in a single numerical quantity. Only ID pairs whose S-T volumes share at least one S-T cell with non-zero weight are considered.

\todo{Talk to and display plots of average precision values.}

The results indicate that SEXTANT with pruning based on device information significantly out-performs the comparison methods for all parameter choices. When using only spatio-temporal information, SEXTANT still out-performs the other methods in all cases except for the most granular setting, when the Jaccard similarity method performs slightly better. For all granularities, SEXTANT with pruning based on device information finds the correct GSM-WiFi pairs for all but at most 10 of the 296,329 devices.

\subsubsection{Impact of Pruning on Runtime}
\label{sec:eval-runtime}

\begin{figure*}[t]
	\centering
	\subfloat{\includegraphics[width=0.45\textwidth]{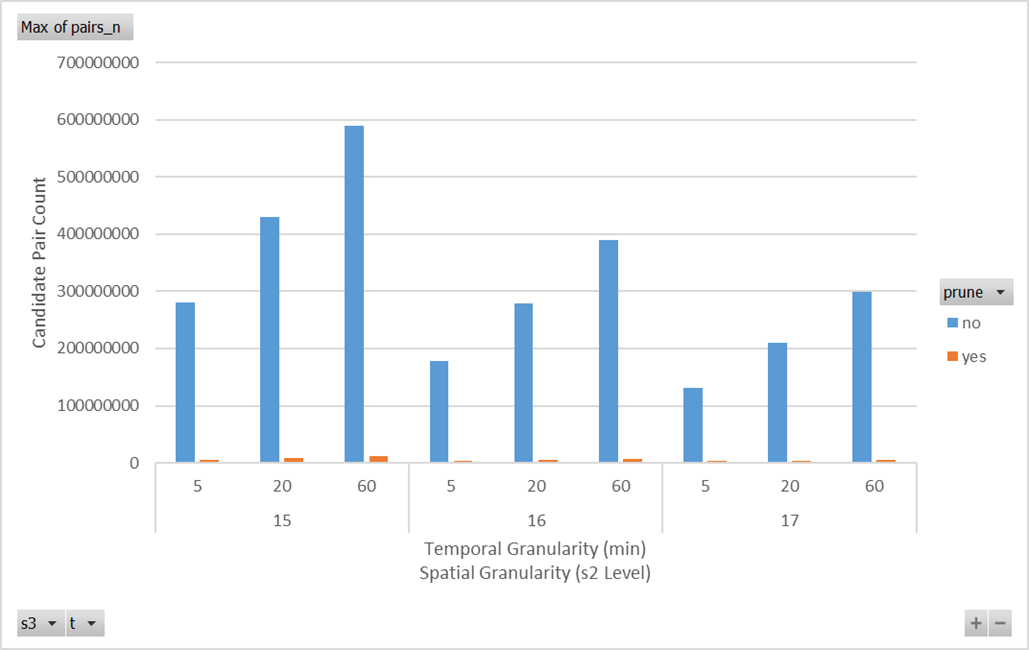}} \quad
	\subfloat{\includegraphics[width=0.45\textwidth]{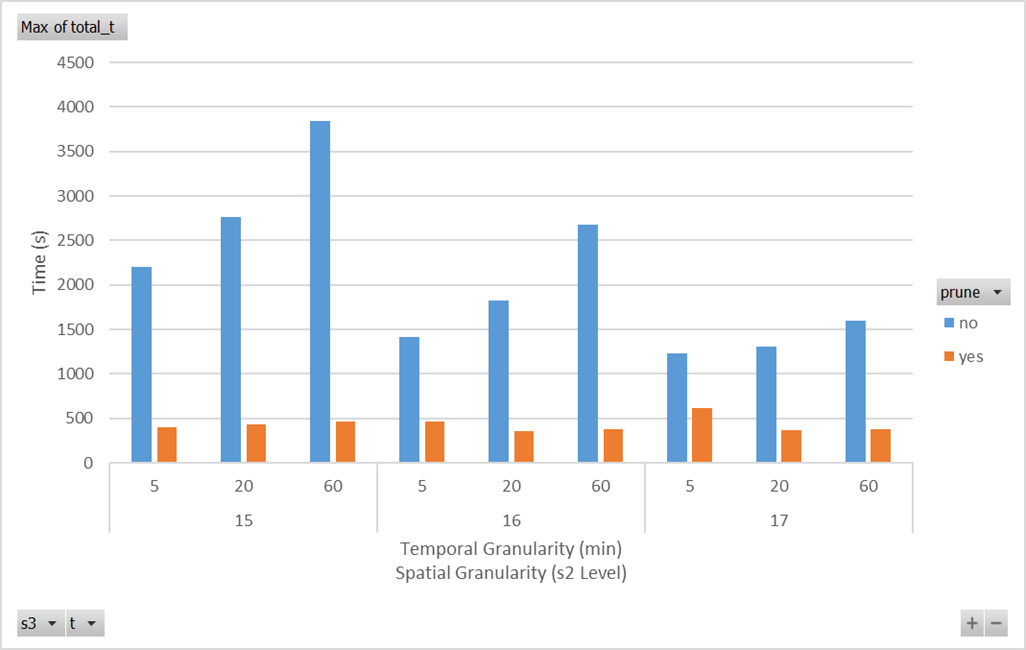}}
	\caption{(a) Number of candidate pairs and (b) runtime efficiency of SEXTANT on the Vegas Events dataset with and without pruning.}
	\label{fig:general-query-runtime}
\end{figure*}

Figure~\ref{fig:general-query-runtime} illustrates the impact of pruning on SEXTANT's runtime efficiency for computing general queries. The results indicate that pruning drastically reduces the runtime, by reducing the number of candidate ID pairs for which the computationally expensive spatio-temporal correlation measures need to be calculated.

%% file: 50-conclusions.tex
\section{Conclusions}
\label{sec:concl}

\subsection{Discussion}
\label{sec:concl-disc}

We have presented SEXTANT, a scalable and accurate method of combining protocol-based pruning and scoring with spatio-temporal trajectory correlation to perform multi-protocol entity resolution. Further, we have demonstrated the surprising efficacy of this combination of techniques at performing multi-protocol resolution on a simulated city-scale dataset, even in the face of significantly reduced spatial and/or temporal resolution. The implication is that post-hoc attempts to address privacy concerns by coarsening or degrading the information made available within one domain (e.g. reducing the spatial resolution of a user’s spatio-temporal trajectory) might be insufficient to obscure sensitive associations when integrated with information available in another (e.g. leaked manufacturer and model information). Privacy must be engineered into the design of the system up front, to reduce the available information as much as possible, and with thorough consideration of the picture that will emerge after integration across domains.


%% file: ms.bbl

%% file: 60-appendix-convexhull.tex
\section{Approximation Algorithm for the Perimeter of the Convex Hull}
\label{apx:proof}

\begin{figure}[t]
	\centering
	\includegraphics[width=0.5\textwidth]{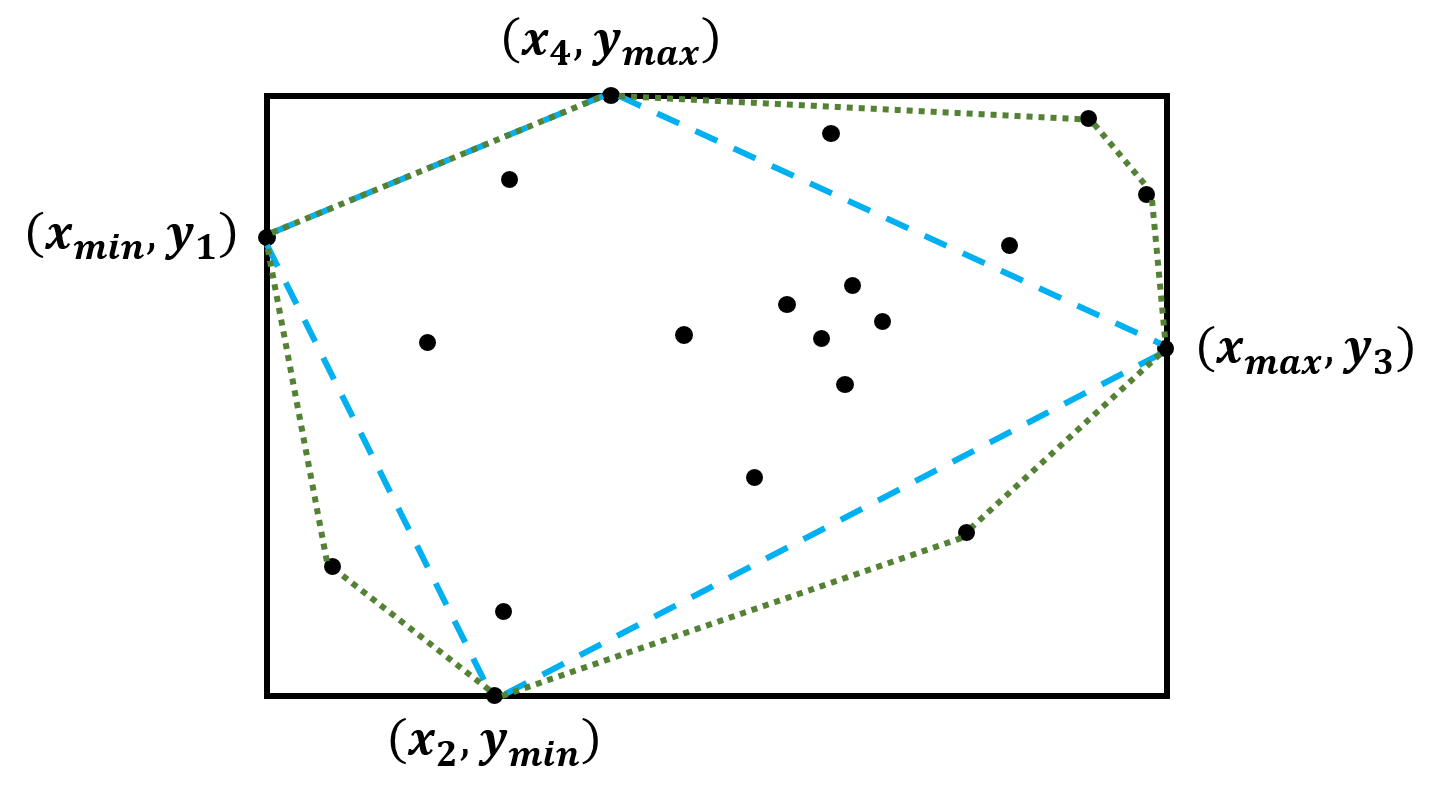}
	\caption{Diagram illustrating the convex hull perimeter approximation algorithm. The black rectangle is the minimal aligned bounding box. The green dotted line is the true convex hull. The blue dashed line is the quadrilateral formed by four points lying along the four sides of the bounding box, respectively.}
	\label{fig:convex-hull}
\end{figure}

The convex hull of a set of $n$ points could, in the worst case, include all $n$ points, and an optimal streaming algorithm for maintaining the convex hull takes $O(\log n)$ per update in the worst case. Thus, to achieve constant space complexity and constant-time updates, approximation is necessary. There exists literature on computing approximate convex hulls, beginning with a 1982 paper by \citet{bentley1982approximation}, but the goal seems to be to achieve better runtime complexity, without necessarily reducing the space complexity, while achieving an approximate convex hull that is very close to the true convex hull. For our context, we want a constant-space algorithm, but we only need to approximate the perimeter of the convex hull, not the convex hull itself.

We now present a simple streaming algorithm to approximate the perimeter of the convex hull of a set of points, that requires only constant space and has constant-time updates, along with a proof that it overestimates the perimeter of the convex hull by at most a factor of $\sqrt{2}$. The intuition is to compute the perimeter of a rectangular bounding box for the set of points, in particular the minimal such rectangle that is aligned with the x and y axes. Although the smallest circumscribed rectangle could be at any angle, the alignment constraint allows for a very simple and efficient algorithm that still provides a reasonable approximation guarantee.

The construction is illustrated in Figure~\ref{fig:convex-hull}. The black rectangle is the minimal aligned bounding box. The green dotted line is the true convex hull. The blue dashed line is the quadrilateral formed by four points lying along the four sides of the bounding box, respectively.

\textbf{Algorithm:} Maintain four values: $x_{min}$ (the minimum x-value seen so far), $x_{max}$ (the maximum x-value seen so far), $y_{min}$ (the minimum y-value seen so far), and $y_{max}$ (the maximum y-value seen so far). Approximate the perimeter of the convex hull as:
\begin{equation}
	2 \cdot (x_{max} - x_{min}) + 2 \cdot (y_{max} - y_{min}).
\end{equation}

\textbf{Theorem:} Given an ordered sequence of spatial coordinates $(x_1, y_1), \ldots, (x_n, y_n)$, let $p$ be the true perimeter of the convex hull, and let $\hat{p}$ be the result from running the streaming convex hull perimeter approximation algorithm above. Then $p \leq \hat{p} \leq \sqrt{2} \cdot p$.

\textbf{Lemma:} For all $a,b \in \R$, $a^2 + b^2 \geq \frac{1}{2} \cdot (a + b)^2$.

\textbf{Proof of Lemma:}
\begin{align*}
	(a - b)^2 \geq 0 & \Rightarrow a^2 - 2ab + b^2 \geq 0 \\
	& \Rightarrow a^2 + b^2 \geq 2ab \\
	& \Rightarrow 2a^2 + 2b^2 \geq a^2 + 2ab + b^2 \\
	& \Rightarrow 2 \cdot (a^2 + b^2) \geq (a + b)^2 \\
	& \Rightarrow a^2 + b^2 \geq \frac{1}{2} \cdot (a + b)^2.
\end{align*}

\textbf{Proof of Theorem:} Consider the rectangle formed by the points $(x_{min}, y_{min})$, $(x_{min}, y_{max})$, $(x_{max}, y_{min})$, and $(x_{max}, y_{max})$. At least one of the given points must lie along each side of the rectangle (equivalently, at least one point must have a coordinate corresponding to each of $x_{min}$, $y_{min}$, $x_{max}$, and $y_{max}$). Without loss of generality, denote them as $(x_{min}, y_1)$, $(x_2, y_{min})$, $(x_{max}, y_3)$, and $(x_4, y_{max})$, respectively. Note that those points must lie within the convex region bounded by the convex hull, so the true perimeter of the convex hull must be at least as large as the perimeter of the quadrilateral formed by those four points. Also, note that each side of that quadrilateral forms the hypotenuse of a right triangle whose other sides lie along the bounding rectangle.  Combining these facts with the above Lemma, we have:
\begin{align*}
	p \geq & \sqrt{(y_1 - y_{min})^2 + (x_2 - x_{min})^2} \\
	& + \sqrt{(x_{max} - x_2)^2 + (y_3 - y_{min})^2} \\
	& + \sqrt{(y_{max} - y_3)^2 + (x_{max} - x_4)^2} \\
	& + \sqrt{(x_4 - x_{min})^2 + (y_{max} - y_1)^2} \\
	\geq & \sqrt{\frac{1}{2} \cdot ((y_1 - y_{min}) + (x_2 - x_{min}))^2} \\
	& + \sqrt{\frac{1}{2} \cdot ((x_{max} - x_2) + (y_3 - y_{min}))^2} \\
	& + \sqrt{\frac{1}{2} \cdot ((y_{max} - y_3) + (x_{max} - x_4))^2} \\
	& + \sqrt{\frac{1}{2} \cdot ((x_4 - x_{min}) + (y_{max} - y_1))^2} \\
	= & \frac{1}{\sqrt{2}} \cdot ((y_1 - y_{min}) + (x_2 - x_{min}) + (x_{max} - x_2) \\
	& + (y_3 - y_{min}) + (y_{max} - y_3) + (x_{max} - x_4) \\
	& + (x_4 - x_{min}) + (y_{max} - y_1)) \\
	= & \frac{1}{\sqrt{2}} \cdot \(2 \cdot (x_{max} - x_{min}) + 2 \cdot (y_{max} - y_{min})\) \\
	= & \frac{1}{\sqrt{2}} \cdot \hat{p} \qquad \Rightarrow \qquad \hat{p} \leq \sqrt{2} \cdot p.
\end{align*}
Furthermore, $p \leq \hat{p}$ because all of the given points lie within the bounding rectangle.  \qed

%% file: 61-appendix-parameters.tex
\section{Parameters Used in Generating the Vegas Events Dataset}
\label{apx:parameters}



\begin{table}[h]
	\centering
	\caption{Probabilistic Parameters}
	\label{table:params}
	\begin{tabular}{l|r|r}
		\toprule
		Parameter				& Signal A & Signal B \\
		\midrule
		Time Between (min) 			& 60  & 45  \\
		Average Location Shift (m)		& 0   & 0   \\
		$\sigma$ of Local Shift (m)	& 10  & 10  \\
		Average Ellipse Major Axis (m)		& 100 & 125 \\
		$\sigma$ of Ellipse Major Axis (m) & 25 & 30 \\
		Average Ellipse Minor Axis (m)		& 50 & 75 \\
		$\sigma$ of Ellipse Major Axis (m) & 10 & 15 \\
		\bottomrule
	\end{tabular}
\end{table}

\todo{List all simulation parameters -- for SUMO, signal generation, and cyber attributes -- either here or in an appendix, for reproducibility. Explain how they express a range of data heterogeneity, etc.}

The SUMO parameter uniformRandomTraffic was set to 0.1 to represent the mobility paths of non-residents.
